\documentclass[11pt,a4paper]{article}
\pdfoutput=1
\usepackage{jheppub}
\usepackage{amsfonts, amssymb, amsmath}
\usepackage{mathrsfs}
\usepackage{graphicx}
\usepackage[utf8]{inputenc}
\usepackage[russian,english]{babel}
\usepackage{enumerate}

\usepackage{color}
\definecolor{dark-gray}{gray}{0.20}
\definecolor{gray}{gray}{0.30}
\definecolor{light-gray}{gray}{0.80}
\definecolor{dark-red}{rgb}{0.7,0,0}
\definecolor{dark-green}{rgb}{0.1,0.4,0}
\definecolor{dark-blue}{rgb}{0.3,0.3,0.7}
\definecolor{light-blue}{rgb}{0.8,0.8,1}
\definecolor{blue}{rgb}{0,0,1}
\definecolor{red}{rgb}{1,0,0}
\definecolor{green}{rgb}{0,1,0}

\usepackage{hyperref}
\hypersetup{
	colorlinks=true,
	linkcolor=dark-blue,
	citecolor=dark-red,
	urlcolor=dark-blue,
	linktoc=page
}

\def\cM{{\cal M}}
\def\cN{{\cal N}}

\def\SO{{\rm SO}}
\def\U{{\rm U}}

\def\i{{\rm i}}

\newcommand{\be}{\begin{equation}}
\newcommand{\ee}{\end{equation}}
\newcommand{\bea}{\begin{eqnarray}}
\newcommand{\eea}{\end{eqnarray}}

\title{Black hole thermodynamics in natural variables: Quadrophenia}

\author{Kiril Hristov}

\affiliation{Faculty of Physics, Sofia University, J. Bourchier Blvd. 5, 1164 Sofia, Bulgaria}
\affiliation{INRNE, Bulgarian Academy of Sciences, Tsarigradsko Chaussee 72, 1784 Sofia, Bulgaria}

\emailAdd{khristov@phys.uni-sofia.bg}

\abstract{
\noindent It was recently observed in \cite{Hristov:2023sxg} for thermal Kerr-Newman black holes in 4d flat space that one can rewrite the conventional thermodynamics on the inner and outer horizons in terms of left- and right-moving variables with a remarkable simplification of the corresponding expressions. With the goal of illustrating the wide applicability of these newly proposed {\it natural} variables, we extend the original observation in four independent directions that can be further superimposed on each other. These four generalizations can be thought of as different deformations of the original 4d Einstein-Maxwell theory, all within the framework of supergravity: higher derivative (HD) corrections in minimal 4d supergravity; additional scalar and vector couplings in matter-coupled 4d supergravity; higher dimensions, in particular 5d minimal supergravity; and a cosmological constant in 4d minimal {\it gauged} supergravity with Anti-de Sitter (AdS) vacuum. Each of these generalizations offers a different lesson about the novel thermodynamics, and we pay special attention to the respective BPS limits that can be understood from fixed point formulae, demonstrating the power of the natural variables to capture the full phase space.
}
\date{\today}
\begin{document}
\maketitle


\section{Introduction and main lessons}
\label{sec:intro}
The laws of black hole thermodynamics were formulated in the 70's by seminal works such as \cite{Bekenstein:1973ur,Bardeen:1973gs,Hawking:1975vcx,Gibbons:1976ue} and have been generalized in numerous directions in the next decades. However, in the lack of a well-understood microscopic picture that a quantum theory of gravity needs to provide, they only remain an analogy to the laws of statistical thermodynamics. String theory so far has been very successful in providing a dual description of the black hole microstates via brane constructions \cite{Strominger:1996sh} and the closely related AdS/CFT correspondence \cite{Maldacena:1997re} in the presence of supersymmetry, but realistic thermal black holes have proven a far more challenging task and exact microscopic calculations are still lacking.
 
Taking into account the above considerations, in the present work we aim to extend the observation of \cite{Hristov:2023sxg} that provided a novel view towards black hole thermodynamics by advocating for a novel set of {\it natural} chemical potentials, suggestive of a simpler microscopic description for the thermal black holes. This construction is based on the bizarre fact that the laws of black hole thermodynamics are actually not unique even for a single spacetime solution. It turns out that one can define a set of chemical potentials and satisfy a corresponding conservation law (known as the first law of black hole thermodynamics) at {\it each} event horizon separately,  \cite{1979NCimB..51..262C}. Perhaps a simple way of understanding this is to consider the original semi-classical calculation of Hawking, \cite{Hawking:1975vcx}, near an event horizon and then to notice that it is equally well applicable both at the inner and the outer horizon of a usual thermal black hole in asymptotically flat 4d spacetime, \cite{Cvetic:1997uw,Cvetic:1997xv,Wu:2004yk} (there are even more event horizons with the same feature in the presence of higher dimensions and/or cosmological constant), see \cite{Cvetic:2018dqf} for a review. Since the first law holds independently for two (or more) independent sets of chemical potentials, it is then possible to take arbitrary linear combinations of the corresponding variables preserving the conservation law. This freedom was explored previously to define the so-called left- and right-moving entropies and temperatures initially in \cite{Cvetic:1997uw,Cvetic:1997xv,Wu:2004yk}, which turned out to be useful for the area product formulae and the Kerr/CFT correspondence, see e.g.\ \cite{Cvetic:2010mn,Castro:2012av,Guica:2008mu,Castro:2010fd,Chen:2012mh,Castro:2013kea} and references thereof. More recently, \cite{Hristov:2023sxg} took the idea a step further by considering the corresponding left- and right-moving on-shell actions. Remarkably, the left- and right-moving on-shell actions turn out to be very simple and fully {\it explicit} functions of the respective chemical potentials, which is not the case for the free energies on the two horizons separately. A summary of this construction for black holes in 4d Einstein-Maxwell theory, embeddable in minimal $\cN=2$ supergravity, can be found in the next section. We recommend the unacquainted reader to browse through the main definitions in  sec.\ \ref{sec:natural}, in particula formulae \eqref{eq:newvar}-\eqref{eq:qsrnew} and \eqref{eq:expl}, before reading about the generalizations below.

As advertised in the abstract, in the present work we extend these results in four independent ways, which we believe illustrate very clearly the wide applicability of the so-called natural chemical potentials. Even though the original construction of \cite{Hristov:2023sxg} does not rely on supersymmetry and can be applied to the black holes in any theory of general relativity coupled to arbitrary matter, it of particular importance from UV perspective~\footnote{In the present work we do not consider any quantum corrections, but it is clear that our results can only be useful once embedded in UV complete theories with {\it finite} corrections.} that the new variables allow for an automatic agreement with the limit to BPS thermodynamics, see \cite{Hristov:2022pmo},~\footnote{See also \cite{Boruch:2023gfn} for closely related results.} which is a very non-trivial feature that has not been observed earlier. In order to pursue this further and emphasize the connection with microscopic entropy counting in string theory, here we focus from the start on gravitational theories exhibiting local supersymmetry, i.e.\ on supergravity. The four different extensions can be then simply seen from a supergravity classification point of view as four ways of deforming minimal 4d $\cN=2$ supergravity.  
\begin{enumerate}[I]
 	\item Higher derivative corrections: we consider four-derivative minimal supergravity in 4d in section \ref{sec:HD}.
	\item Scalars and additional matter: we consider additional vector multiplets in the so-called 4d STU model in section \ref{sec:scalars}.
	\item Higher dimensions: we consider black holes in 5d minimal supergravity in section \ref{sec:5d}.
	\item Cosmological constant: we consider asymptotically AdS$_4$ black holes with spherical and higher-genus horizon topologies in minimal {\it gauged} supergravity in section \ref{sec:ads}.
\end{enumerate}
These four generalizations can be superimposed on each other since the supergravity framework in principle allows for an infinite set of matter couplings, which string theory helps to constrain. We believe that the chosen four directions are representative of all the main features one can expect in more complicated settings within supergravity. In each of these cases we pay special attention to the BPS limit, which can be derived in several alternative ways, allowing us for a direct comparison with literature, see \cite{Ooguri:2004zv,Cassani:2019mms,Hosseini:2019iad,Hristov:2021qsw}. Let us note that the proposal of \cite{Hristov:2023sxg} that we extend here aims at the suggestive reformulation of the on-shell action in terms of the natural chemical potentials that cover the {\it complete} black hole phase space. This is in contrast with much of the other recent progress, focused on particular simplifying limits such as the near-BPS limit, the near-extremal limit of low temperature, or the opposite limit of high-temperature expansion. Due to the prohibitively immense amount of literature on these special limits, the related thermal CFT results, and the ``extended'' thermodynamics in AdS, see e.g.\ \cite{Kastor:2009wy,Cvetic:2010jb,DiPietro:2014bca,Maldacena:2016upp,Larsen:2019oll,Iliesiu:2020qvm,Heydeman:2020hhw,Benjamin:2023qsc} and references therein and thereof, we do not attempt here to relate our present results to any of these interesting developments, only discussing the BPS limit and the general thermal phase space whenever we are able to present fully explicit results.

\subsubsection*{Main lessons}
Given that each of the listed theories and pertaining solutions exhibit their own technical differences and particularities, which we explain in the relevant sections, the present work contains a considerable number of new results that cannot be simply summarized in an introductory fashion. Here we can focus on the main lessons for the new construction of natural variables that each of these generalizations offers, in hope of clarifying the main underlying principles. In the same order of appearance, our main conclusions are the following.
\begin{enumerate}[I]
 	\item As already remarked in \cite{Hristov:2023sxg}, the split of left- and right-moving sectors seems in close analogy to the split of holomorphic and anti-holomorphic variables for the BTZ black hole and the related thermodynamics of two-dimensional conformal field theories (CFT$_2$'s), see e.g.\ \cite{Kraus:2005zm} and \cite{Detournay:2012ug}. At the level of 4d supergravity, there is a natural split in holomorphic, or F-term (chiral superspace), and real, or D-term (full superspace), supersymmetric invariants that in turn influence the black hole on-shell action, see e.g.\ \cite{Lauria:2020rhc} for a review of the superconformal formalism. Considering two particular F-term four derivative corrections to the two derivative theory, we show  in section \ref{sec:HD} that only the left-moving sector of the black hole thermodynamics gets appropriately corrected, whereas the right-moving sector does not feel the corrections at all, see \eqref{eq:4dlrsectors} and the discussion around it. Even though this needs to be proven for a much larger set of theories and black hole solutions, this result is already very suggestive that the main structure and appeal of the left- and right-moving sectors, taking simple form in terms of the corresponding chemical potentials, is preserved with the inclusion of higher derivative corrections. This is to be contrasted with the area product formulas of \cite{Cvetic:2010mn,Castro:2012av} that were shown in \cite{Castro:2013pqa,Detournay:2012ug} to generically lose their mass-independence in the presence of higher derivatives, ultimately leading this research direction to lose momentum. If we instead push the present results further and prove that the left- and right-moving structure is preserved with arbitrary higher derivatives, it would be clear that the natural variables indeed play a crucial role in understanding the perturbative corrections in supergravity and string theory.

	\item Our results in the case of matter-coupled 4d supergravity in section \ref{sec:scalars} serve the purpose of firmly establishing the main features of the left- and right-moving sectors also in presence of additional scalars and electromagnetic charges. As we show there, the main structure of black hole thermodynamics in minimal supergravity remains virtually intact, with the important caveat that the explicit expressions in terms of the natural variables become increasingly more complicated. Every additional vector multiplet brings an extra electric chemical potential, which unfortunately means that we are not able to present fully analytic expressions for the left- and right-moving on-shell actions in the general STU model. Nevertheless, due to the explicit calculations in the attached {\it Mathematica} file, we are able to prove in general the simplification of a {\it static} left-moving sector as in \cite{Hristov:2023sxg}, and we analytically reproduce the expected BPS limit as in \cite{Hristov:2022pmo}, in agreement with the OSV formula, \cite{Ooguri:2004zv}. Additionally, we show that the $X^0 X^1$ truncation to only a single vector multiplet allows once again for a full analytic expression of the left-moving sector, exemplifying how the original expressions in \cite{Hristov:2023sxg} get generalized.

	\item Working out the example of 5d asymptotically flat black holes with S$^3$ topology reveals clearly two new interesting features. The first one is that the left- and right-moving sectors in 5d exhibit a symmetry upon the flip of sign of the electric charge and the exchange of the two angular momenta, see \eqref{eq:expl5d} (the 5d Lorentz group allows for two independent rotations, and each of the two sectors only feels one of them, a clear generalization of the 4d case). Given that some of the matter-coupled 4d models can also be derived from a 5d compactification, such a hidden symmetry is automatically incorporated there, but is much more difficult to see from a 4d perspective. This apparent further simplification in 5d minimal supergravity also leads to the interesting question of whether the natural variable structure shows particular simplification for models coming from string compactifications in comparison with generic other models in supergravity (4d minimal ungauged supergravity is notoriously {\it} not by itself embeddable in string theory, unlike the cubic models coming from 5d). The second feature, which is also shared with matter-coupled 4d black holes, is the existence of the so-called {\it almost} BPS bound, \cite{Goldstein:2008fq,Bena:2009ev}, which is in close analogy to the BPS bound except that it leads to a vanishing left-moving on-shell action and non-vanishing right-moving one. From the 4d perspective this bound corresponds to the {\it underrotating}, or {\it slow-rotating}, branch of black holes that actually exhibit a hidden supersymmetry, \cite{Hristov:2012nu,Hristov:2014eba}, and a rather interesting higher derivative structure, see \cite{Hristov:2021qsw}. The existence of such an almost BPS bound is in fact a generic feature of many black holes and the natural variables can thus be potentially used as a tool for searching for these bounds. Given these additional features, the 5d example hints at an important {\it practical} use of the natural variables for technical simplification of calculations, independent of the underlying question whether they have a truly fundamental meaning.

	\item The addition of cosmological constant presents the following additional puzzle for the choice of {\it natural} variables. Asymptotically AdS$_4$ black holes generically exhibit {\it four} distinct event horizons that independently allow for the definition of chemical potentials and respective first law of black hole thermodynamics, even though only two of them can possibly be real for the thermal solutions. There are thus various ways of generalizing the natural variable construction. We explore several options, such as defining separate left- and right-moving variables for each different pair of horizons, paying particular attention to the two real horizons, see \eqref{eq:pairwise}. We also define four particular linear combinations of {\it all} four horizons, which instead of left- and right-moving we label with the cardinal directions $W$, $N$, $E$, and $S$, see \eqref{eq:directional}. An additional feature of AdS asymptotics is also the existence of a much more general set of compact horizon topologies, encompassing Riemann surfaces $\Sigma_\frak{g}$ of arbitrary genus, see \cite{Caldarelli:1998hg}. It turns out that the explicit expressions for the chemical potentials are prohibitively complicated and even numerically it is very hard to gain much useful intuition on the behavior of the respective on-shell actions. Nevertheless, we are able to make several interesting observations that point to the ``directional'' variables as the most promising ones for useful simplifications. Perhaps the most tantalizing hint in this direction is the fact that the sum of all on-shell actions, $I_{(i)}$, is purely topological:
\be
	I_{(1)} + I_{(2)} + I_{(3)} + I_{(4)} = 2 \pi\, \kappa\, L^2\, |1-\frak{g}|\ ,
\ee
where the subscript $(i)$ labels the on-shell action at the respective horizon, $L$ is the AdS$_4$ scale, $\frak{g}$ is the genus of the Riemann surface (we exclude the toroidal case $\frak{g} =1$ from the analysis), and $\kappa = \pm 1$ for spherical and hyperbolic horizon curvature, respectively. The above rule holds for the thermal AdS$_4$ black holes with arbitrary mass, electomagnetic charges and rotation (which is allowed when $\frak{g} = 0$), and is therefore strikingly simple.~\footnote{Note that similarly simple expressions were already noticed in \cite{Xu:2013zpa,Xu:2014qaa} for the sum of the entropies on all horizons. The simplification ultimately stems from Vieta's formula applied to the roots of quartic polynomials, see below.
It can similarly be shown that the sum of the on-shell actions remains a purely topological quantity also in presence of four derivative corrections via \cite{Bobev:2020egg,Bobev:2021oku}.} It would be interesting to understand its holographic/microscopic origin. We furthermore consider the two topologically distinct BPS limits: twisted higher-genus black holes, reproducing the recent result of \cite{BenettiGenolini:2023ucp} via the natural variable picture; and untwisted rotating black holes, where we find some simplification for the ``directional'' on-shell actions, but are only able to partially relate to the results of \cite{Choi:2018fdc,Cassani:2019mms} due to the insufficient power of our numerical analysis.

\end{enumerate}

\subsubsection*{Reading guide}
The rest of the paper is organized as follows. In section \ref{sec:natural} one can find the main definitions of the natural variables, a summary of the main calculations and results of \cite{Hristov:2023sxg} and an extended discussion on the BPS limit. These main features are in many ways shared by all the extensions that follow in the next four sections, which can be read entirely independently of each other. Their individual contents were enlisted above, and at the end of each of these sections we give comments and outlook on how to connect the corresponding results to the results of the other sections.  We therefore recommend the reader to start with the next section and then pick freely from the rest of the sections depending on their specific interest. All explicit calculations have been performed using {\it Mathematica}, and we have included the relevant {\it .nb} file with the present submission, organized section by section in the same order as the paper.

\section{Natural variables and the BPS limit}
\label{sec:natural}
Let us start by introducing the logic of the so-called natural chemical potentials, which can be applied to any thermal black hole solution. We limit ourselves here to considering exactly two instances of an event horizon for a single spacetime solution, but the logic can be straightforwardly brought to the case of more horizons, which we explicitly consider in section \ref{sec:ads}. A general comment is in order on the working {\it meaning} of an event horizon we use here: we assume the horizons appear at the radial positions of {\it all} roots of the expression for the $g^{rr}$ component of the metric ($r$ being the radial coordinate). Some, or all, of these roots may be complex, in turn making the corresponding chemical potentials also complex, which is perfectly acceptable in the present formalism. The only condition that we enforce in order to calculate the inverse temperature $\beta$ at each horizon is the smooth capping off of the Euclideanized geometry at the position of the respective horizon (the Euclidean time coordinate gets periodically identified $\tau \sim \tau + \beta$). Importantly, it is not possible to ensure the smooth capping off of the geometry and thus a global regularity at {\it all} horizons at the same time, which eventually leads to independent thermodynamic laws at each horizon separately. 

Let us be more concrete and look at two such horizons, appearing at positions $r_-$ and $r_+$ (due to the previous comments we do {\it not} assume an ordering relation such as $r_- \leq r_+$). Each stationary black hole spacetime asymptoting to a maximally symmetric vacuum will have a set of conserved asymptotic charges: the energy (or mass in units of $c=1$) $M$ and a set of independent angular momenta $J_\alpha$, whose number depends on the spacetime dimensions. Additionally we allow for a set of abelian Maxwell fields with corresponding conserved electric charges $Q_i$. All these charges are computed asymptotically and therefore do not depend on the number of black hole horizons. Conversely, the corresponding conjugate variables, or chemical potentials, $\beta_\pm, \Omega^\alpha_\pm, \Phi^i_\pm$ are quantities computed at each horizon independently. Their subscripts therefore denote the belonging to the particular horizon. It is then possible to verify on a case by case basis for all black hole horizons that the first law of thermodynamics holds:
\be
\label{eq:firstlaw}
	\beta_\pm\, \delta M = \delta S_\pm + \beta_\pm \Omega^\alpha_\pm\, \delta J_\alpha + \beta_\pm \Phi^i_\pm\, \delta Q_i \ .
\ee

Furthermore, the above quantities always obey an additional relation, called the Smarr formula, \cite{Smarr:1972kt}. It will not play an important role in what follows as it remains the same in the new variables without additional simplifications, but we discuss it explicitly for our first example of Kerr-Newman black holes in Einstein-Maxwell theory below. We can also generically compute the corresponding on-shell action of the spacetime using either the boundary conditions $\{\tau \in [0, \beta_- ), r \in [r_-, \infty) \}$, or $\{\tau \in [0, \beta_+ ), r \in [r_+, \infty) \}$, producing $I_-$ and $I_+$ respectively. The full calculation of the on-shell action and its asymptotic regularization naturally depends on the asymptotic vacuum, so we implicitly always include the corresponding Gibbons-Hawking-York (GHY) boundary terms,  \cite{Gibbons:1976ue,York:1986it},  relevant either for flat space or for asymptotic AdS. The final result for the on-shell actions always satisfies the so-called {\it quantum statistical relation}, \cite{Gibbons:1976ue,Gibbons:2004ai},
\be
\label{eq:qsr}
	I_\pm (\beta_\pm, \Omega_\pm, \Phi_\pm) = \beta_{\pm} M - S_\pm - \beta_\pm \Omega^\alpha_{\pm} J_\alpha - \beta_{\pm} \Phi^i_{\pm} Q_i\ ,
\ee
such that the first law of thermodynamics equates to the extremization of the on-shell action, $\delta I_\pm = 0$. This in turn allows us to define formally the thermal partition function as the Euclideanized path integral with the boundary conditions fixed by the asymptotic charges and the GHY terms. Here we continue the discussion at a purely classical level.

At this stage we notice that the two independent laws of thermodynamics and quantum statistical relations can be freely mixed with each other if we take the same linear combinations for all chemical potentials and on-shell actions. The following choices for left- and right-moving variables have already been considered in \cite{Cvetic:1997uw,Cvetic:1997xv,Cvetic:2010mn,Castro:2012av} and references thereof,
\bea
\label{eq:newvar}
\begin{split}
	\beta_{l,r} :=& \frac12\, (\beta_+ \pm \beta_-)\ , \qquad \qquad \qquad \omega_{l,r}^\alpha :=  \frac12\, (\beta_+ \Omega^\alpha_+ \pm \beta_- \Omega^\alpha_-)\ , \\  \varphi^i_{l,r} :=& \frac12\, (\beta_+ \Phi^i_+ \pm \beta_- \Phi^i_-)\ , \qquad \qquad S_{l,r} :=  \frac12\, (S_+ \pm S_-)\ ,
\end{split}
\eea
leading to an alternative version of the first law,
\be
\label{eq:lrfirstlaw}
	\beta_{l,r}\, \delta M = \delta S_{l,r} + \omega_{l,r}^\alpha\, \delta J_\alpha + \varphi^i_{l,r}\, \delta Q_i\ .
\ee
Reference \cite{Hristov:2023sxg} then took one step further and defined also
\be
\label{eq:newvar2}
	I_{l,r} := \frac12\, ( I_+ \pm I_-)\ ,
\ee
with corresponding quantum statistical relation,
\be
\label{eq:qsrnew}
	I_{l,r} (\beta_{l,r}, \omega_{l,r}, \varphi_{l,r}) = \beta_{l,r} M - S_{l,r} - \omega^\alpha_{l,r} J_\alpha - \varphi^i_{l,r} Q_i\ .
\ee
Remarkably, the expressions one finds for $I_l$ and $I_r$ in terms of the corresponding left- and right-moving chemical potentials are much simpler and in fact {\it analytic}, contrary to the case for $I_+$ and $I_-$.

\subsection{The BPS limit and a fixed point formula}
Generically, the BPS limit is always defined by a linear relation on the asymptotic charges of the type
\be
	M^\text{BPS} = a^\alpha\, J_\alpha + b^i\, Q_i\ ,
\ee
with the coefficients $a^\alpha$ and $b^i$ particular constants that depend on the details of both the theory and the black hole solution in question. This relation in turn enforces a similar linear relation (or even two such relations for the asymptotically flat solutions) for the chemical potentials. For the sake of keeping the notation simple, let us focus solely on the left-moving sector, where we generically encounter the following linear relation between chemical potentials
\be
	\beta_l^\text{BPS} = c_\beta\, \omega^\beta_l + d_j\, \varphi^j_l\ ,
\ee
again with constant coefficients $c_\beta$ and $d_j$ depending on the theory and solution type. Note that the analogous relation in the right-moving sector typically also features an additive imaginary multiple of $\pi$, see later. At the level of the left-invariant on-shell action, imposing the above linear constraints leads to the following non-conventional expression for the quantum statistical relation,
\be
	I_l^\text{BPS} = - S_l + ( ( c_\beta\, \omega^\beta_l + d_j\, \varphi^j_l)\, a^\alpha - \omega^\alpha_l)\, J_\alpha + (( c_\beta\, \omega^\beta_l + d_j\, \varphi^j_l)\, b^i - \varphi^i_l)\, Q_i\ .
\ee
The {\it canonical} conjugate variables to the corresponding charges are actually defined by the requirement that
\be
	I_{l}= - S_{l,r} - \omega^\alpha_\text{BPS}\, J_\alpha - \varphi^i_\text{BPS}\, Q_i\ ,
\ee
such that we need to perform the redefinition
\be
	\omega^\alpha_\text{BPS} := \omega^\alpha_l - ( c_\beta\, \omega^\beta_l + d_j\, \varphi^j_l)\, a^\alpha\ , \qquad \varphi^i_\text{BPS} := \varphi^i_l - ( c_\beta\, \omega^\beta_l + d_j\, \varphi^j_l)\, b^i\ .
\ee
We therefore see that what we called natural variables do not identically match the canonical BPS chemical potentials, but are a simple linear combination of them. This is reminiscent in spirit, but distinctly different in explicit realization, to the procedure proposed in \cite{Cabo-Bizet:2018ehj,Cassani:2019mms} specifically for the BPS limit of the general thermodynamics that focuses only on a single black hole horizon (the outer-most one in the special cases that the radii are real).

Going to the right-moving sector, after using the corresponding linear relations for the chemical potentials, we remarkably retrieve 
\be
	 I^\text{BPS}_r = 0\, \qquad \Rightarrow\, \qquad I^\text{BPS}_+ = I^\text{BPS}_l\ ,
\ee
with the latter relation allowing us to make the explicit comparison of our results in the BPS limit and the results obtained via the prescription of \cite{Cabo-Bizet:2018ehj,Cassani:2019mms}. We have no first principles proof that $I_r$ is automatically vanishing, but this is indeed the case in all the following examples with some interesting twists on this in the cases of Mink$_5$ and AdS$_4$ asymptotics. In the former case we also find an alternative BPS limit where the places of the $I_l$ and $I_r$ are swapped, while in the latter case we encounter four different horizons and in turn four independent sectors, only one of which becomes identically zero in the BPS limit.

Another important comparison with our results in the BPS limit is provided by the direct supersymmetric calculation of the entropy and on-shell action using only the strict supersymmetric black hole solutions, which can be argued to follow from a simple fixed point formula, \cite{Hristov:2021qsw}. In the asymptotically flat cases this formula is known as the OSV formula, \cite{Ooguri:2004zv}, and was more recently derived from the full thermal solutions in \cite{Hristov:2022pmo}. For asymptotically AdS spacetimes the formula generalizes to many more examples and is known as gluing of gravitational blocks, \cite{Hosseini:2019iad}. In all of these cases the relevant fixed point formulae depend on simple input: the holomorphic prepotential $F (X)$ that defines the underlying supergravity. We give more details pertaining to each separate case below.

\subsection{Summary of \cite{Hristov:2023sxg}: 4d Einstein-Maxwell}
In 4d Einstein-Maxwell theory, in the conventions of \cite{Hristov:2023sxg} ($G_N = 1$), the action is given by
\be
	 I_\text{EM} = \frac1{16 \pi} \int_{\cal M}  {\rm d}^4 x\, \sqrt{-g}\, \left( R - \frac14 F_{\mu \nu} F^{\mu \nu} \right)\ ,
\ee
with an abelian field strength $F = {\rm d} A$ and mostly positive metric signature. This is the bosonic part of the action of 4d minimal {\it ungauged} supergravity, which can be uniquely defined via the prepotential
\be
	F_\text{min} (X) = - 2 i\, X^2\ ,
\ee
with $X$ the holomorphic section, which in this case can be thought of as an auxiliary field that only serves as an intermediate step of writing the theory, see e.g.\ \cite{Lauria:2020rhc} for a review on the full superconformal formalism.
We have the most general stationary black hole solution~\footnote{We fix a typo in \cite{Hristov:2023sxg}, correcting the power of $\sin \theta$ in the $e^0$ vierbein.}
\be
	{\rm d} s^2 = - \frac{\Delta (r)}{\rho^2} \left( {\rm d} t - a \sin^2 \theta\, {\rm d} \phi \right)^2 + \frac{\rho^2}{\Delta (r)}\, {\rm d} r^2 
+ \rho^2\, {\rm d} \theta^2 +  \frac{\sin^2 \theta}{\rho^2} \left( a\, {\rm d} t - (r^2+a^2)\, {\rm d} \phi \right)^2\ ,
\ee
with 
\be
	\Delta(r) = r^2 - 2 M r + a^2 + Q^2 + P^2\ , \, \qquad \rho^2 = r^2 + a^2 \cos^2 \theta\ ,
\ee
and background gauge field
\be
	A = - \frac{Q\, r}{\rho^2}\,  \left( {\rm d} t - a \sin^2 \theta\, {\rm d} \phi \right) -  \frac{P\, \cos \theta}{\rho^2}\,   \left( a\, {\rm d} t - (r^2+a^2)\, {\rm d} \phi \right)\ ,
\ee
leading to the conserved electromagnetic charges $Q$ and $P$. The full solution is completely specified by $Q, P$, the mass (or energy) $M$ and the angular momentum $J = a M$. The two roots of the function $\Delta (r)$ are given by
\be
\label{eq:radii}
	r_{\pm} = M \pm \sqrt{M^2 - a^2 - Q^2 - P^2}\ ,
\ee
and correspond to the positions of the outer and the inner event horizons.

Following the standard definitions, see \cite{York:1986it,Braden:1990hw}, the corresponding chemical potentials are thus
\bea
\label{eq:betaEM}
\begin{split}
	\beta_\pm & = 2 \pi \frac{r_\pm^2 + a^2}{r_\pm - M}\ , \qquad \quad \Omega_\pm = \frac{a}{r_\pm^2 + a^2}\ , \\
	\Phi_\pm &= \frac{Q\, r_\pm}{r_\pm^2 + a^2}\ , \qquad \qquad \Psi_\pm = \frac{P\, r_\pm}{r_\pm^2 + a^2}\ ,
\end{split}
\eea
and the Bekenstein-Hawking entropies of the two horizons are given by
\be
\label{eq:entropy}
	S_\pm = \pi (r_\pm^2 + a^2)\ .
\ee

With these definitions it is straightforward to verify the first law,
\be
	\beta_\pm\, \delta M = \delta S_\pm + \beta_\pm \Omega_\pm\, \delta J + \beta_\pm \Phi_\pm\, \delta Q + \beta_\pm \Psi_\pm\, \delta P \ ,
\ee
as well as the Smarr relation, \cite{Smarr:1972kt},
\be
	\beta_\pm\, M = 2\, S_\pm + 2\, \beta_\pm \Omega_\pm\,  J + \beta_\pm \Phi_\pm\,  Q + \beta_\pm \Psi_\pm\,  P \ .
\ee

Let us for simplicity set $P=0$ (due to electromagnetic duality $P$ and its conjugate $\Psi_\pm$ can always be uniquely destroyed) and perform the change of variables to left- and right-moving chemical potentials, following \eqref{eq:newvar}. We then immediately find 
\be
	\omega_l = 0\ ,
\ee
signifying that the left-moving sector is actually static. The remaining chemical potentials are all non-trivial, but their explicit form becomes rather simple, as explicitly shown in \cite{Hristov:2023sxg}. The first law and Smarr relation take the following forms in terms of the new variables, respectively,
\be
	\beta_l\, \delta M = \delta S_l + \varphi_l\, \delta Q\ , \qquad \beta_r\, \delta M = \delta S_r + \omega_r\, \delta J + \varphi_r\, \delta Q\ ,
\ee
and
\be
	\beta_l\, M = 2\, S_l + \varphi_l\,  Q\ , \qquad \beta_r\, M = 2\, S_r + 2\, \omega_r\, J + \varphi_r\,  Q\ .
\ee
Most remarkably, we can derive the following expressions for the on-shell actions in the two sectors, 
\bea
\begin{split}
\label{eq:expl}
		I_l (\beta_l, \varphi_l) = & \frac1{8 \pi}\, (\beta_l^2 - 2 \varphi_l^2)\ , \\
	I_r (\beta_r, \omega_r, \varphi_r) = \frac1{16}\,  & \left( 3 \beta_r -  \sqrt{\beta_r^2 + 8 \varphi_r^2} \right)^{3/2} \,  \sqrt{\frac{\beta_r + \sqrt{\beta_r^2 + 8 \varphi_r^2}}{4 \pi^2 + \omega_r^2}} \ ,
\end{split}
\eea
that carry the full {\it classical} thermodynamic information of the general Kerr-Newman class of black holes.

\subsubsection*{Conjugate variables and stability}

For the newly defined on-shell actions $I_{l,r}$, we find
\be
		\frac{\partial I_{l,r}}{\partial \beta_{l,r}} = M \ , \qquad \frac{\partial I_{r}}{\partial \omega_{r}} = - J\ , \qquad \frac{\partial I_{l,r}}{\partial \varphi_{l,r}} = - Q\ ,
\ee
justifying the claim that the conjugate variables are precisely $\omega, \varphi$ in \eqref{eq:qsrnew} and elsewhere. Note that the above relations can also be used as a derivation of the asymptotic charges given the explicit form of the on-shell actions.

If we take the left- and right-moving sectors proposed above as serious contenders for fundamental thermodynamic description, we should be also able to address more involved questions such as the stability of these solutions against thermal perturbations. We can define the corresponding heat capacities,
\be
\label{eq:heatcap}
	C_{l,r} : = - \beta_{l,r}\, \frac{\partial S_{l,r}}{\partial \beta_{l,r}}\ ,
\ee
which can be easily calculated from \eqref{eq:expl} and the additional identities
\be
\label{eq:StoI}
	 S_l (\beta_l, \varphi_l) = I_l (\beta_l, \varphi_l) \ , \qquad \left(1 + \frac{\omega_r^2}{4 \pi^2} \right)\, S_r (\beta_r, \omega_r, \varphi_r) = I_r (\beta_r, \omega_r, \varphi_r) \ ,
\ee
We thus find
\be
	C_l = - \frac{\beta_l^2}{4 \pi}\ , \qquad C_r = - \pi^2 \beta_r\, \sqrt{\frac{(3 \beta_r- \sqrt{\beta_r^2 + 8 \varphi_r^2}\, ) (\beta_r+ \sqrt{\beta_r^2 + 8 \varphi_r^2}\, )}{(4 \pi^2 + \omega_r^2)^3}}\ , 
\ee
such that both heat capacities are manifestly negative in the space where physical black holes are expected, $\beta_l > 0, \beta_r > \varphi_r$. This suggests that the black holes in consideration are unstable against thermal fluctuations. This is precisely coincident with the analogous conclusion based on the usual specific heat calculations at the outer horizon, see e.g.\ \cite{Avramov:2023eif}.

\subsubsection*{The BPS limit and the OSV formula}
In the BPS limit, in the absence of magnetic charge, we simply find $M = \pm Q$, depending on the sign of the electric charge, see \cite{Hristov:2022pmo} for a more extensive discussion of the BPS limit. The thermodynamic potentials in turn can be shown to satisfy,  \cite{Hristov:2023sxg},
\be
\label{eq:BPSchem}
	\beta_l = \pm 2\, \varphi_l\ ,\qquad \omega_r = (s)\, 2 \pi \i\ , \qquad \beta_r = \pm \varphi_r\ ,
\ee
where only the first and third equalities are related to the sign of the electric charge, while the imaginary right-moving angular velocity can take both signs, $s = \pm 1$, independently. This is due to the BPS identity at the inner and outer horizons, which reads 
\be
	\beta_\pm\, \Omega_\pm  = \pm\, (s)\, 2 \pi i\ .
\ee
Note that the sign ambiguity fundamentally stems from the definition of the $r_\pm$, which only assumes $r_+ > r_-$ for the cases where these quantities are real. In the BPS limit they are actually each other's complex conjugate, not allowing a simple ordering relation.
We further find identically vanishing right-moving on-shell action and simplified left-moving one,
	\be
	I^\text{BPS}_l  =  \frac{\varphi_l^2 }{4 \pi}\ , \quad \quad \quad I^\text{BPS}_r = 0\ ,
\ee
as anticipated. Further, due to the fact that $M$ and its conjugate $\beta_l$ are fixed in terms of $Q$ and $\varphi_l$, respectively, the first law of BPS thermodynamics becomes
\be
	\delta I^\text{BPS}_l = - \delta S_l + \varphi_l\, \delta Q\ ,
\ee
and the corresponding quantum statistical relation is
\be
	I^\text{BPS}_l = -  S_l + \varphi_l\, Q\ ,
\ee
with the opposite than usual sign in front of $\varphi_l$. We can restore the {\it canonical} normalization of the BPS chemical potentials by the simple identification
\be
	\varphi_\text{BPS} := - \varphi_l\ ,
\ee
such that
\be
	I^\text{BPS}_l = -  S_l - \varphi_\text{BPS}\, Q = \frac{\varphi_\text{BPS}^2 }{4 \pi}\ .
\ee
We have thus formally reproduced the exact same result as in \cite{Hristov:2022pmo}, but notably in a rather different way that includes considering the thermodynamics of both horizons that do {\it not} coincide in the BPS limit.

This is in agreement with the OSV formula, \cite{Ooguri:2004zv}, which is given by summing over two (in the case of vanishing magnetic charges, identical) copies of the prepotential,
\be
	I^\text{OSV} = i\, (F_\text{min} - \bar F_\text{min}) \Big|_{(X = \frac{\varphi_\text{BPS}}{4 \sqrt{\pi}})} = \frac{\varphi_\text{BPS}^2 }{4 \pi}\ ,
\ee
see again \cite{Hristov:2022pmo} for an extended discussion.

\section{Higher derivatives}
\label{sec:HD}
Let us now consider the simplest higher derivative deformation of the Einstein-Maxwell theory in 4d, allowed within the framework of supergravity. This corresponds to adding a couple of four derivative invariants, \cite{Bergshoeff:1980is,Butter:2013lta} that we call the $\mathbb{W}$ and $\mathbb{T}$ invariant, respectively. Their couplings to the original two derivative theory are fully fixed upto two arbitrary coefficients $c_1$ and $c_2$, which should be considered small compared to the Newton constant, $c_{1,2} << 1$. This situation was considered in \cite{Castro:2013pqa,Charles:2016wjs} and revisited in \cite{Bobev:2020egg,Bobev:2021oku} including cosmological constant, and the details of the formalism have been discussed in details in these references. For our purposes here we can just write down the final bosonic part of the resulting four derivative action, featuring again the metric $g_{\mu\nu}$ and gauge field $A_\mu$,
\be
	I_{4 \partial} = I_\text{EM} + c_1\, I_\mathbb{W} + c_2\, I_\mathbb{T}\ ,
\ee
where
\be 
	I_\mathbb{T} = \int_{\cal M}  {\rm d}^4 x\, \sqrt{-g}\, \left( \frac{2}{3} R^2 - 8 (\nabla_\mu F^{- \mu\nu}) (\nabla^\rho F^+_{\rho \nu}) - 2 (R_{\mu\nu} + F^-_{\mu\rho} F^{+ \rho}_\nu) \right)\ ,
\ee
\be
	I_\mathbb{W} = I_\text{GB} - I_\mathbb{T}\ , \qquad I_\text{GB} = \int_{\cal M}  {\rm d}^4 x\, \sqrt{-g}\, \left( R^{\mu\nu\rho\sigma} R_{\mu\nu\rho\sigma} - 4 R^{\mu\nu} R_{\mu\nu} + R^2 \right)\ ,
\ee
with $I_\text{GB}$ the Gauss-Bonnet invariant, which is purely topological and does not change local physics but still plays an important role in what follows. The above Lagrangian follows from augmenting the two-derivative prepotential with additional four derivative terms,
\be
	F_{4 \partial} (X; A_\mathbb{W}, A_\mathbb{T}) = - 2 i\, X^2 - 2 i\, c_1\, A_\mathbb{W} - 2 i\, c_2\, A_\mathbb{T}\ , 
\ee
where $A_\mathbb{W, T}$ are auxiliary composite scalars that introduce the additional terms as given above, see e.g.\ \cite{Bobev:2021oku,Hristov:2021qsw}. Note that both of these HD terms are F-terms that come from chiral superspace integrals, i.e.\ they can be encoded in the prepotential as above.

The task of correcting the Kerr-Newman black hole thermodynamics discussed in the previous section is greatly facilitated by the observation of \cite{Charles:2016wjs} (see also \cite{Castro:2013pqa}) that $I_\mathbb{T}$ is a sum of terms that corresponds to squares of the original Einstein-Maxwell field equations, which means the original solutions also exactly solve the equations of motion of the new theory. Furthermore, the entire contribution of $I_\mathbb{T}$ on the solutions vanishes identically. The only non-trivial contribution is then entirely topological, as the GB term evaluates on any black hole solution to
\be
	I_\text{GB} = 32 \pi^2\, \chi(\cM) = 64 \pi^2\ ,
\ee
since spherical topology fixes completely the Euler characteristic, $\chi (\text{KN}) = 2$. Being a boundary term, the GB action does not change the asymptotic charges and chemical potentials, and therefore all relations \eqref{eq:betaEM} between $M, J, Q$ and $\beta_\pm, \omega_\pm, \varphi_\pm$ on the two horizons remain precisely as in the two-derivative theory.  However, it changes the entropies and the on-shell actions,
\be
	S_\pm^{4 \partial} = S_\pm^{2 \partial} - 64 \pi^2 c_1\ , \qquad  I_\pm^{4 \partial} = I_\pm^{2 \partial} + 64 \pi^2 c_1\ ,
\ee
such that the first law of thermodynamics and quantum statistical relation remain valid in a trivial way, given that the additional term is a constant that does not depend on the chemical potentials. The fact that the quantities get shifted by the same constant factor on both horizons leads to the interesting result that only the left-moving sector actually sees the HD correction at all, while the right-moving sector is effectively identical with the two derivative one,
\be
\label{eq:4dlrsectors}
	I^{4 \partial}_l =  I_l^{2 \partial} + 64 \pi^2 c_1\ , \qquad I^{4 \partial}_r =  I_r^{2 \partial}\ .
\ee 
Explicitly, we again find that the left-moving sector is effectively static, and given by
\be
	I^{4 \partial}_l (\beta_l, \varphi_l) = \frac1{8 \pi}\, (\beta_l^2 - 2 \varphi_l^2) + 64 \pi^2 c_1\ , 
\ee
while the right-moving sector is again given by \eqref{eq:expl}. Note a remarkable feature of \eqref{eq:4dlrsectors}: it tells us that the HD F-terms only have an effect on the left-moving sector and do not correct the right-moving sector. On the other hand, it is clear from the superconformal formalism and the way they only appear in the prepotential vs. K\"ahler potential, that F-terms correspond to holomorphic corrections while D-terms (coming from full superspace integrals) correspond to real corrections. Although we have not proven this in full generality for an arbitrary matter-coupled theory, it is tempting to speculate that \eqref{eq:4dlrsectors} is just another manifestation of the above fact. As already emphasized in the introductory discussion, it is interesting to note that the higher derivatives considered above already spoil the nice structure of the area product formula, see \cite{Castro:2013pqa}. The fact that these corrections preserve the structure of the on-shell actions, as clear from above, is reassuring that there is a more fundamental appeal in the present construction.

\subsection*{The BPS limit and the OSV formula}

Given the minimal change in thermodynamic potentials, it is easy to derive the BPS limit, which again corresponds to $M = \pm Q$ and fixes the chemical potentials as in \eqref{eq:BPSchem}. The on-shell actions are thus
	\be
	I^\text{BPS}_{4 \partial, l}  =  \frac{\varphi_l^2 }{4 \pi} + 64 \pi^2 c_1\ , \quad \quad \quad I^\text{BPS}_{4 \partial, r} = 0\ .
\ee
and one can again define the canonical chemical potential $\varphi_\text{BPS}$ such that
\be
	I^\text{BPS}_{4 \partial, l} = -  S^{4 \partial}_l - \varphi_\text{BPS}\, Q = \frac{\varphi_\text{BPS}^2 }{4 \pi} + 64 \pi^2 c_1 \ .
\ee
Once again this is in exact accordance with the OSV formula, \cite{Ooguri:2004zv}, which in this case is given by, \cite{Hristov:2021qsw},
\be
	I^\text{OSV}_{4 \partial} =  i\, (F_{4 \partial}  - \bar F_{4 \partial} ) \Big|_{(X = \frac{\varphi_\text{BPS}}{4 \sqrt{\pi}}; A_\mathbb{W} = 16 \pi^2, A_\mathbb{T} = 0)} = \frac{\varphi_\text{BPS}^2 }{4 \pi} + 64 \pi^2 c_1 \ ,
\ee
as can be shown directly in the superconformal formalism, \cite{LopesCardoso:1998tkj,LopesCardoso:1999fsj}.

\subsubsection*{Relation to the other extensions}
Some of the HD corrections in the superconformal formalism have been relatively well understood also in presence of additional matter couplings, additional dimensions and/or cosmological constant, see \cite{deWit:2010za,Banerjee:2011ts,Butter:2013lta,Gold:2023ymc} and references thereof. Each of these further deformations however introduces specific complications that are outside the present scope.~\footnote{Note that here we have been discussing single centered thermal black holes, but one can also consider the question of HD corrections to multi-centered BPS black holes, see e.g.\ \cite{Hu:2023qhs} for a recent discussion.} We should note that the simplest task, which was in fact already accomplished in \cite{Bobev:2020egg,Bobev:2021oku}, is to consider the same four derivative couplings in minimal {\it gauged} supergravity, allowing the straightforward superposition of the results of this section with those of section \ref{sec:ads}. This is no longer true in case of matter-coupled or higher-dimensional supergravity due to the fact that the original two derivative solutions are no longer solutions to the corrected theory. In these cases there has only been a substantial progress in understanding the BPS solutions via the supersymmetry variations, \cite{LopesCardoso:1998tkj,LopesCardoso:1999fsj,Baggio:2014hua,Hristov:2016vbm,Hristov:2021qsw,Bobev:2022bjm,Cassani:2022lrk}, while the general thermal black holes are much harder to analyze, see \cite{Cassani:2022lrk} for partial results.

\section{4d STU model}
\label{sec:scalars}
The general matter-coupled 4d supergravity theory has the following form, following the notation of \cite{Chow:2014cca},
\be
	 I_\text{sugra} = \frac1{16 \pi} \int_{\cal M}  {\rm d}^4 x\, \Big[ \sqrt{-g}\, \left( R - \frac12\, f_{AB} \partial_\mu \Phi^A \partial^\mu \Phi^B - \frac14\, k_{IJ} F^I_{\mu \nu} F^{J, \mu \nu} \right) + \frac{\epsilon^{\mu\nu\rho\sigma}}4\, h_{IJ} F^I_{\mu\nu} F^J_{\rho \sigma} \Big]\ ,
\ee
where $f_{AB}, k_{IJ}, h_{IJ}$ are functions of the real scalars $\Phi^A$.~\footnote{Note that for the purposes of presenting the black hole solutions more succinctly, we present the Lagrangian in a rather unconventional way. In the usual supergravity conventions, see \cite{Andrianopoli:1996cm}, each separate vector multiplet comes with a $\U(1)$ gauge field $A^i$ together with a complex scalar $z^i$.} For the STU model there are three complex scalars, $z^i = \chi_i + i\, e^{-\varphi_i}$, $i=1, 2, 3$, and four gauge fields $A^I$, $I = 0, 1, 2, 3$,  such that we have $\Phi^A = (\varphi_1, \varphi_2, \varphi_3, \chi_1, \chi_2, \chi_3)$ and the corresponding metric $f_{AB} = \text{diag}  (1, 1, 1, e^{2 \varphi_1}, e^{2 \varphi_2}, e^{2 \varphi_3})$. The matrices that correspond to $k_{IJ}$ and $h_{IJ}$ are non-diagonal and can be read off from \cite{Chow:2014cca}. In the usual supergravity formalism, this theory is uniquely defined via the corresponding prepotential,
\be
	F_\text{STU} (X) = - 2 i\, \sqrt{X^0 X^1 X^2 X^3}\ .
\ee

The solutions are carefully spelled out in the original references, \cite{Chow:2013tia,Chow:2014cca}, and here we can illustrate the main features. The general solutions are labeled by ten parameters $m$, $a$, $\delta_{0, 1, 2, 3}$, $\gamma_{0, 1, 2, 3}$ that uniquely determine the conserved asymptotic charges $M$, $J$, $Q_{0, 1, 2, 3}$, $P^{0, 1, 2, 3}$. For simplicity we look at the purely electric case, $P^I = 0$, corresponding to  $\gamma_0= \gamma_1 = \gamma_2 = \gamma_3 = 0$. The metric is given by
\be
\label{eq:minmetric}
	{\rm d} s^2 = - \frac{R - U}{W} ({\rm d} t+\omega_3)^2 + W \left( \frac{{\rm d} r^2}{R} + \frac{{\rm d} u^2}{U} + \frac{R U}{a^2 (R-U)}\, {\rm d} \phi^2 \right)\ ,
\ee
with
\be
	R(r) = r^2 - 2 m r +a^2\ , \qquad U(u) = a^2 - u^2\ ,
\ee
\be
	W^2 = (R - U)^2 + L^2 + 2 (R-U) (2 M r + V)\ , \qquad \omega_3 = \frac{U L}{a (R-U)}\, {\rm d} \phi\ ,
\ee
with
\be
	L (r) = \frac{m}{4 \prod_I e^{\delta_I}}\, \Big[ m \prod_I (e^{2 \delta_I} -1) + \left(  \sum_I e^{2 \delta_I} + \sum_{I < J < K} e^{2 (\delta_I + \delta_J + \delta_K)} \right)\, r \Big]\ ,
\ee
and $V(u)$ a similar linear function, see \cite{Chow:2014cca}. Above, we already assumed that the total NUT charge is vanishing. The usual angular coordinate $\theta$ from the previous sections is simply recovered by the change of variables $u = a\, \cos \theta$.

The physical mass and angular momentum are given by
\be
	M = \frac{m}4\, \sum_I \cosh (2 \delta_I)\ , \qquad J = \frac{a m}8\, \Big[  \sum_I e^{2 \delta_I} + \sum_{I < J < K} e^{2 (\delta_I + \delta_J + \delta_K)}  \Big]\ .
\ee

The solution for the gauge field $A$ can be compactly written as
\be
\label{eq:gaugefield}
	A = - W \frac{\partial}{\partial \delta} \left( \frac{{\rm d} t + \omega_3}{W} \right)\ ,
\ee
leading to the conserved electric charges
\be
\label{eq:Qcharges}
	Q_I = \frac{m}4\, \sinh (2 \delta_I)\ .
\ee
The explicit form of the scalars is more convoluted, see again \cite{Chow:2014cca}.

Looking at the $g^{rr}$ component of the metric, the two roots of the function $R (r)$ are given by
\be
	r_\pm = m \pm \sqrt{m^2 - a^2}\ ,
\ee
fixing the position of the two horizons. In turn we use these expressions to evaluate the chemical potentials and corresponding Bekenstein-Hawking entropies,
\bea
\label{eq:beta}
\begin{split}
	\beta_\pm & = 2 \pi \frac{L(r_\pm)}{r_\pm - M}\ , \quad \quad \Omega_\pm = \frac{a}{L (r_\pm)}\ , \\
	\Phi^I_\pm &= \frac1{L (r_\pm)}\, \frac{\partial L (r_\pm)}{\partial \delta_I} \ , \quad \quad S_\pm = \pi\,  L (r_\pm)\ .
\end{split}
\eea
The explicit expressions for the chemical potentials in this case depend generically on six free parameters parametrizing the solution, and they can be found in full detail in the auxiliary {\it Mathematica} file. Using these expressions we were able to show that the first law of black hole thermodynamics, see \eqref{eq:firstlaw}, indeed holds as expected. 

In the general STU case it turns out that the expressions depend on too many variables, and even the change to left- and right-moving variables is not enough to ensure finding simple analytic expressions for the on-shell actions. Only in the case of the so-called $X^0 X^1$ truncation, discussed below as a separate subsection, we were able to find the left-moving on-shell action explicitly. For the remainder of this section we list the general analytic results for the STU model that we could simplify in {\it Mathematica}, most notably presenting fully the BPS limit.

One of the important analytic results we can prove in full generality for the STU model, see again the attached {\it .nb} file, is that the left-moving sector is once again {\it static}, i.e.\
\be
\omega_l = 0\, ,
\ee 
without any further assumptions. Additionally, we again observe the following relations between the entropies and the on-shell actions in the two sectors:
\be
	 S_l (\beta_l, \varphi^I_l) = I_l (\beta_l, \varphi^I_l) \ , \qquad \left(1 + \frac{\omega_r^2}{4 \pi^2} \right)\, S_r (\beta_r, \omega_r, \varphi^I_r) = I_r (\beta_r, \omega_r, \varphi^I_r) \ ,
\ee
in exact agreement with \eqref{eq:StoI}. Unfortunately, we were unable to find the exact expressions for $I_{l,r}$ in terms of $\beta_{l,r}, \omega_r, \varphi^I_{l,r}$ in an analytic way.

\section*{The BPS limit and the OSV formula}
In the BPS limit we were able to fully simplify the expressions for the on-shell actions, exemplifying that the procedure leads to the expected answer. Based on the abstract supersymmetry algebra, see \cite{Hristov:2022pmo} for the detailed discussion, we can derive the following supersymmetric limit of the black holes in the STU model
\be
	M = \pm \sum_{I = 0}^3\, Q_I\ .
\ee
Fixing the mass in such a way in turn leads to
\be
	2\, \beta_l = \pm \sum_{I = 0}^3\, \varphi^I_l\ , \qquad \omega_r = (s)\, 2 \pi i\, \qquad 4\, \beta_r = \pm \sum_{I = 0}^3\, \varphi_r^I\ ,
\ee
which is a clear generalization of the minimal case. We again have $s = \pm 1$ due to the ambiguity in the ordering of $r_\pm$, and
\be
	\beta_\pm\, \Omega_\pm  = \pm\, (s)\, 2 \pi i\ .
\ee
Using the above constraints, we explicitly derive,~\footnote{See again the auxiliary {\it .nb} file for the detailed calculation.}
\be
	I_l^\text{BPS} = \frac1{4 \pi}\, \sqrt{\prod_I (\varphi_l^I - \frac12\, \sum_J \varphi_l^J)} \ , \qquad I_r^\text{BPS} = 0\ .
\ee
The quantum statistical relation is given in this case by
\be
	I_l^\text{BPS} = -S_l - (\varphi_l^I - \frac12\, \sum_J \varphi_l^J)\, Q_I\ ,
\ee
and we need to define the canonical conjugate variables in the BPS limit,
\be
	\varphi^I_\text{BPS} := \varphi_l^I - \frac12\, \sum_{J = 0}^3 \varphi_l^J\ .
\ee
In terms of these variables we find
\be
\label{eq:BPSSTU}
	I_l^\text{BPS} = -S_l - \varphi^I_\text{BPS}\, Q_I = \frac1{4 \pi}\, \sqrt{\varphi^0_\text{BPS} \varphi^1_\text{BPS} \varphi^2_\text{BPS} \varphi^3_\text{BPS}}\ .
\ee
This is precisely the expected answer from the OSV formula in the absence of magnetic charges, see again \cite{Hristov:2022pmo}:
 \be
	I^\text{OSV}_\text{STU} = i\, (F_\text{STU} - \bar F_\text{STU}) \Big|_{(X^I = \frac{\varphi^I_\text{BPS}}{4 \sqrt{\pi}})} =\frac1{4 \pi}\, \sqrt{\varphi^0_\text{BPS} \varphi^1_\text{BPS} \varphi^2_\text{BPS} \varphi^3_\text{BPS}}\ .
\ee

\subsection{The $X^0 X^1$ truncation}
This truncation is attained by the pair-wise identification
\be
	X^2 = X^0\ , \qquad X^3 = X^1\ , 
\ee
such that the truncated action is defined by the simpler prepotential,
\be
	F_{X^0 X^1} = - 2 i\, X^0 X^1\ .
\ee
In this truncation, which has an automatic embedding in $\cN=8$ supergravity via the STU model, also corresponds to a truncation of $\cN = 4$ supergravity. In this case we were able to analytically find the left-moving on-shell action using the attached {\it .nb} file:
\be
	I_l = \frac{\beta_l^2 - (\varphi^1_l)^2 - (\varphi_l^2)^2}{8 \pi}\ ,
\ee
together with $\omega_l = 0$. Using this explicit form, we are also able to verify~\footnote{Note that the electric charges as defined in \eqref{eq:Qcharges} are rescaled with a factor of $4$ with respect to the minimal supergravity case, which is due to differing conventions.}
\be
		\frac{\partial I_{l}}{\partial \beta_{l}} = M \ ,  \qquad \frac{\partial I_{l}}{\partial \varphi^0_{l}} = -2\, Q_0\ ,  \qquad \frac{\partial I_{l}}{\partial \varphi^1_{l}} = -2\, Q_1\ ,
\ee
as expected. We also find the left-moving heat capacity, given by
\be
	C_l = - \frac{\beta_l^2}{4 \pi}\ ,
\ee
precisely as in the minimal case.

We can again revisit the BPS limit for completeness, even if the results have already been presented above for the general STU model. The constraint between charges gives $M = \pm (Q_0 + Q_1)$, and we find
\be
	I_l^\text{BPS} = \frac{\varphi_l^0 \varphi_l^1}{4 \pi}\ , \qquad I_r^\text{BPS} = 0\ ,
\ee
with a quantum statistical relation
\be
	I_l^\text{BPS} = -S_l + \varphi_l^1\, Q_0 +  \varphi_l^0\, Q_1\ . 
\ee
We again need a canonical redefinition of the form
\be
	\varphi^0_\text{BPS} := - \varphi^1_l\ , \qquad \varphi^1_\text{BPS} := - \varphi^0_l\ ,
\ee
such that
\be
	I_l^\text{BPS} = -S_l - \varphi_\text{BPS}^0\, Q_0 -  \varphi_\text{BPS}^1\, Q_1 = \frac{\varphi_\text{BPS}^0 \varphi_\text{BPS}^1}{4 \pi}\ . 
\ee
Again, this is in precise agreement with the OSV formula, and coincides with the $X^0 X^1$ truncation of \eqref{eq:BPSSTU}, as expected.

\subsubsection*{Relation to the other extensions}
From the results above it is evident that the basic structure of the left- and right-moving sectors of minimal supergravity persists in the matter-coupled theory, but explicit calculations and expressions become increasingly more unwieldy with the addition of more vector multiplets. We expect a similar situation upon superimposing the extra matter couplings with the other possible deformations, which we already commented upon for the case of HD corrections. It is clear from the 5d calculations in the next section that the additional couplings to matter will be similarly computationally (but not conceptually) more cumbersome, and we expect the same to be true in the asymptotically AdS case as well. Note also that, just like the case discussed next, 4d matter-coupled black holes exhibit both a BPS and an extremal non-BPS limit that is known as {\it almost} BPS, see below. The latter limit is also expected to yield a vanishing $I_l$ and non-vanishing $I_r$, but the computation is technically more involved and is therefore only presented in the minimal 5d case below.

\section{5d Einstein-Maxwell}
\label{sec:5d}
In 5d Einstein-Maxwell theory, in the conventions of \cite{Chong:2005hr} (setting $G_N = 1$),
\be
	 I_\text{5d} = \frac1{16 \pi} \int_{\cal M}  {\rm d}^5 x\, \Big[ \sqrt{-g}\, \left( R - \frac14 F_{\mu \nu} F^{\mu \nu} \right) + \frac{\varepsilon^{\mu\nu\rho\sigma\tau}}{12 \sqrt{3}}\, A_\mu F_{\nu \rho} F_{\sigma \tau} \Big]\ ,
\ee
with an abelian field strength $F = {\rm d} A$ and mostly positive metric signature. This is the bosonic part of the action of 5d minimal {\it ungauged} supergravity. Adopting a 4d point of view,~\footnote{To uniformize notation with the rest of the sections, we can schematically think of the Kaluza-Klein compactification of 5d supergravity on a circle. The explicit map can be found in \cite{Gaiotto:2005gf,Behrndt:2005he}, but here we only need the resulting prepotential in order to discuss the BPS limit below. One can ignore this subtlety for the main calculations of this section, performed directly in 5d on the black hole solutions of interest.} the theory corresponds to the prepotential
\be
\label{eq:5dprepot}
	F_{5 d} (X) = - \frac{(X^1)^3}{X^0}\ ,
\ee
with two holomorphic sections $X^{0,1}$. This is due to the fact that upon compactification, there appears an additional Kaluza-Klein vector multiplet in 4d.

Here we make the choice of considering the most general Kerr-Newman type of solutions, see \cite{Cvetic:1996xz,Chong:2005hr},~\footnote{Unlike 4d, in higher dimensions there can be completely disjoint classes of solutions that have the same asymptotic charges. In 5d ungauged supergravity there are two such classes: the black holes with S$^3$ topology and black rings with S$^1 \times$S$^2$ topology. The most general black string, which can be further compactified on its extended direction to a ring, can be found in \cite{Compere:2010fm}.} given by the metric
\bea
\begin{split}
	{\rm d} s^2  =  &- \frac{{\rm d} t}{\rho^2} \left( \rho^2\, {\rm d} t + 2 q\, \nu \right)  + \frac{2 q\, \nu\, \sigma}{\rho^2}+ \frac{2 m \rho^2-q^2}{\rho^4}\, ({\rm d} t - \sigma)^2 + \frac{\rho^2}{\Delta (r)}\, {\rm d} r^2 \\
& + \rho^2\, {\rm d} \theta^2 +   (r^2+a^2)\, \sin^2 \theta\, {\rm d} \phi^2 +   (r^2+b^2)\, \cos^2 \theta\, {\rm d} \psi^2\ ,
\end{split}
\eea
with 
\bea
\begin{split}
	\nu &= b\, \sin^2 \theta\, {\rm d} \phi + a\, \cos^2 \theta\, {\rm d} \psi\ , \quad \qquad  \quad \sigma = a\, \sin^2 \theta\, {\rm d} \phi + b\, \cos^2 \theta\, {\rm d} \psi \ , \\
	\Delta(r) &= \frac{(r^2+a^2) (r^2+b^2) + q^2+2 a b q}{r^2} - 2 m\ , \qquad \rho^2 = r^2 + a^2 \cos^2 \theta + b^2 \sin^2 \theta\ ,
\end{split}
\eea
and background gauge field
\be
	A = - \frac{\sqrt{3}\, q}{\rho^2}\,  \left( {\rm d} t - \sigma \right)\ .
\ee
The full solution is completely specified by the parameters $m, q, a, b$, which in turn fix the asymptotic charges
\bea
\begin{split}
	M & = \frac{3 \pi m}{4}\ , \qquad \qquad Q = \frac{\sqrt{3} \pi q}{4}\ , \\
	J_a & = \frac{\pi}{4}\, (2 a m+ b q)\ , \quad \quad J_b = \frac{\pi}{4}\, (2 b m+ a q)\ ,
\end{split}
\eea
with the two angular momenta being the Cartan subgroup of the full 5d rotation group, $\SO(4)$. We can also define for later convenience another couple of linearly independent angular momenta,
\be
\label{eq:5dnewJ}
	J_x := \frac12\, (J_a + J_y)\ , \qquad \qquad J_y := \frac12\, (J_a - J_b)\ .
\ee

The two positive roots of the function $\Delta (r)$ are given by
\be
\label{eq:radii}
	r_{\pm} = \frac1{\sqrt{2}}\, \sqrt{2 m -a^2-b^2 \pm \sqrt{(2 m -a^2-b^2)^2 - 4 (a b+q)^2}}\ ,
\ee
and correspond to the positions of the outer and the inner event horizons. Note that all metric functions actually depend on the square of the radial coordinate, so we strictly speaking get a doubling of these roots, the other two simply corresponding to taking an overall negative sign. Since the on-shell actions and corresponding temperatures and entropies pick up the same overall sign, the other two roots do not really carry additional information and are automatically linearly dependent. We can therefore simply ignore the negative signs and repeat the construction of left- and right-moving variables with the two roots above, exactly as in the 4d cases.

The corresponding chemical potentials were also calculated in \cite{Cvetic:1996xz,Chong:2005hr},
\bea
\label{eq:beta}
\begin{split}
	\beta_\pm & = 2 \pi r_\pm\, \frac{(r_\pm^2 + a^2) (r_\pm^2+b^2) + a b q}{r^2_\pm - (a b+q)^2}\ , \quad \quad \Phi_\pm = \frac{\sqrt{3}\, q\, r_\pm^2}{(r_\pm^2 + a^2) (r_\pm^2+b^2) + a b q}\ , \\
	\Omega^a_\pm &= \frac{a\, (r_\pm^2+b^2)+ b q}{(r_\pm^2 + a^2) (r_\pm^2+b^2) + a b q}\ , \quad \quad \quad \Omega^b_\pm = \frac{b\, (r_\pm^2+a^2)+ a q}{(r_\pm^2 + a^2) (r_\pm^2+b^2) + a b q}\ ,
\end{split}
\eea
and the Bekenstein-Hawking entropies are given by
\be
\label{eq:entropy}
	S_\pm = \frac{\pi^2}{2 r_\pm}\, ((r_\pm^2 + a^2) (r_\pm^2+b^2) + a b q)\ .
\ee
The above set of conserved charges and conjugate potentials at each horizon can be shown to satisfy the first law of black hole thermodynamics, see again the attached {\it .ng} file.

Considering now the left- and right-moving variables, we follow the standard redefinitions from \eqref{eq:newvar}, in addition also defining a new couple of angular velocities,
\be
	\omega^x_{l,r} := \frac12\, (\omega^a_{l,r} + \omega^b_{l,r})\ , \qquad \qquad \omega^y_{l,r} := \frac12\, (\omega^a_{l,r} - \omega^b_{l,r})\ ,
\ee
conjugate to the angular momenta defined in \eqref{eq:5dnewJ}, respectively. This choice of variables proves particularly useful in the explicit calculations, since we find
\be
	\omega^x_l = 0\ , \quad \omega^y_l \neq 0\ , \quad \qquad \omega^x_r \neq 0\ , \quad \omega^y_r = 0\ .
\ee
It follows that each of the two sectors feels only one of the angular momenta, an interesting generalization of the 4d picture that now exhibits an exact symmetry between left- and right-moving variables. Furthermore, the parameters characterizing the black hole solution can be explicitly inverted in terms of the new variables:
\be
	m = \frac{3 \beta_l^2 - \varphi_l^2}{6\, ((\omega^y_l)^2+\pi^2)}\ , \quad q = \frac{(\sqrt{3} \beta_l + \varphi_l)\, \varphi_l}{3\, ((\omega^y_l)^2+\pi^2)}\ , \quad (a-b) =  \frac{(\sqrt{3} \beta_l + \varphi_l)\, \omega^y_l}{\sqrt{3}\, ((\omega^y_l)^2+\pi^2)}\ ,
\ee
and
\be
	m = \frac{3 \beta_r^2 - \varphi_r^2}{6\, ((\omega^x_r)^2+\pi^2)}\ , \quad q = \frac{(\sqrt{3} \beta_r - \varphi_r)\, \varphi_r}{3\, ((\omega^x_r)^2+\pi^2)}\ , \quad (a+b) =  \frac{(\sqrt{3} \beta_r - \varphi_r)\, \omega^x_r}{\sqrt{3}\, ((\omega^x_r)^2+\pi^2)}\ .
\ee
As already manifest by these expressions, we find a completely symmetric answer for the on-shell actions, given by
\bea
\begin{split}
\label{eq:expl5d}
		I_l (\beta_l,  \omega^y_l, \varphi_l) & = \frac{\pi\, (\sqrt{3} \beta_l + \varphi_l)^2\, (\sqrt{3} \beta_l - 2 \varphi_l)}{24 \sqrt{3}\, ((\omega^y_l)^2+\pi^2)}\ , \\
	I_r (\beta_r, \omega^x_r, \varphi_r) & = \frac{\pi\, (\sqrt{3} \beta_r - \varphi_r)^2\,  (\sqrt{3} \beta_r + 2 \varphi_r) }{24 \sqrt{3}\, ((\omega^x_r)^2+\pi^2)}\ .
\end{split}
\eea
Upon a brief inspection, the two sectors map to each other via exchange of angular momenta and a flip of the sign for the electric charge.

\subsubsection*{Conjugate variables and stability}

Having explicitly solved for the two sectors, it is straightforward to verify the expected conjugate relations
\be
		\frac{\partial I_{l,r}}{\partial \beta_{l,r}} = M \ , \quad \frac{\partial I_{l}}{\partial \omega^y_{l}} = - J_y\ , \quad \frac{\partial I_{r}}{\partial \omega^x_{r}} = - J_x\ , \quad \frac{\partial I_{l,r}}{\partial \varphi_{l,r}} = - Q\ ,
\ee
which can also be taken as definitions of the conserved charges.

We also find the following relations between the corresponding entropies and the on-shell actions of the two sectors
\be
	\frac{\pi^2 + (\omega_l^y)^2}{2 \pi^2}\, S_l (\beta_l, \omega_l^y, \varphi_l) = I_l (\beta_l, \omega_l^y, \varphi_l) \ , \quad \frac{\pi^2 + (\omega_r^x)^2}{2 \pi^2}\, S_r (\beta_r, \omega_r, \varphi_r) = I_r (\beta_r, \omega_r, \varphi_r) \ ,
\ee
which allows us to determine the heat capacities, defined again via \eqref{eq:heatcap}. We find
\be
	C_l = - \frac{\pi^3\, \beta_l\, (3 \beta_l^2 - \varphi_l^2)}{8\, ((\omega^y_l)^2+\pi^2)^2 }\ , \qquad C_r = - \frac{\pi^3\, \beta_r\, (3 \beta_r^2 - \varphi_r^2)}{8\, ((\omega^x_r)^2+\pi^2)^2 }\ , 
\ee
such that both heat capacities are manifestly negative in the space where physical black holes, again in agreement with our 4d results.

\subsection{Two (almost) BPS limits}

We have seen the apparent symmetry between the left- and right-moving sectors in the general thermal case upon a flip of the electric charge and the exchange of angular momenta, which is a new feature in comparison to the 4d case. When considering the BPS limit it also leads to the logical question of whether we have an alternative limit with vanishing left-moving sector and non-vanishing right-moving one. And indeed this is the case, as we can consider two {\it inequivalent} BPS limits, corresponding to $M = \sqrt{3}\, Q$ and $M = - \sqrt{3}\, Q$. Remarkably, the two cases actually have a drastically different supersymmetry properties and physical interpretation.

Let us first choose the positive sign, requiring $Q > 0$, which corresponds to the standard BPS limit that ensures supersymmetry is preserved. We find the following relations between the chemical potentials,
\be
	\beta_l = \sqrt{3}\, \varphi_l\ , \qquad \omega^x_r = \pm  i \pi\ , \qquad \beta_r = \frac1{\sqrt{3}}\, \varphi_r\ ,
\ee
leading in turn to
\be
	I_l^\text{BPS} = \frac{2 \pi\, \varphi_l^3}{3 \sqrt{3}\,  ((\omega^y_l)^2+\pi^2) }  \ , \qquad \qquad I_r^\text{BPS} = 0\ .
\ee
We see that this situation is in close analogy to the BPS limit in 4d, with a vanishing right-moving on-shell action. For the left-moving sector, we have the quantum statistical relation
\be
	I_l^\text{BPS} = - S_l + 2\, \varphi_l\, Q - \omega_l^y\, J_y\ .
\ee
In order to restore the canonical normalization, we need to define
\be
	\varphi_\text{BPS} := -2\, \varphi_l\ , \qquad \omega^y_\text{BPS}:= \omega^y_l
\ee
such that
\be
\label{eq:5dbps}
	I_l^\text{BPS} = - S_l - \varphi_\text{BPS}\, Q - \omega_\text{BPS}^y\, J_y = - \frac{\pi\, \varphi_\text{BPS}^3}{12 \sqrt{3}\,  ((\omega^y_\text{BPS})^2+\pi^2) } \ .
\ee
The result is again expected from the OSV formula, \cite{Ooguri:2004zv}, which in this case requires more careful discussion. We already presented the prepotential that the 5d theory exhibits from a four-dimensional point of view, see \eqref{eq:5dprepot}. In order to apply the OSV formula, we further need to translate the conserved black hole charges to the 4d perspective. The electric charge $Q$ simply becomes (upto a $\sqrt{3}$ prefactor due to different conventions) the electric charge $Q^1$ carried by the gauge field $A^1$, conjugate to the chemical potential $\varphi^1_\text{BPS}$. On the other hand, the Kaluza-Klein gauge field $A^0$ actually carries both an electric charge, which precisely corresponds to the 5d angular momentum $J^y$, and a magnetic charge $P^0 = 1$. This is precisely the topological charge carried by the fibration of the circle over S$^2$ that results in the three-sphere from a 5d perspective. The OSV formula therefore produces
\be
	I^\text{OSV}_{5d} = i\, (F_{5d} - \bar F_{5d}) \Big|_{(X^0 = \varphi^0_\text{BPS} + i \pi P^0; X^1 = \varphi^1_\text{BPS})} =  - \frac{2 \pi\, (\varphi^1_\text{BPS})^3}{ ((\varphi^0_\text{BPS})^2+\pi^2) }\ ,
\ee
in agreement with \eqref{eq:5dbps} upon the identification $\varphi^0 = \omega^y$ and $\varphi^1 = \varphi/ (2 \sqrt{3})$ in the 4d and 5d pictures of the same physical system.

Let us now focus on the case $Q < 0$ and $M = - \sqrt{3}\, Q$, which corresponds to the so-called {\it almost} BPS limit, \cite{Goldstein:2008fq,Bena:2009ev}. It is a peculiar case where boundary conditions on the locally existing Killing spinors are twisted such that globally they remain undefined in the usual ungauged supergravity, even if the underlying 5d solution is clearly analogous to the BPS one. It was later shown that supersymmetry is restored in the so-called {\it flat} gauged supergravity, which retains the same bosonic sector but exhibits charged gravitini, see \cite{Hristov:2012nu,Hristov:2014eba}. Interestingly, these solutions are rather different from a 4d point of view, where they fall in the class of {\it underrotating}, or {\it slow-rotating}, extremal black holes, see also \cite{LopesCardoso:2007qid,Gimon:2007mh,Bossard:2012xsa,Chow:2014cca} and references thereof.
On the level of the chemical potentials, the almost BPS limit (abbreviated aBPS below) is analogous to the BPS one, upon exchanging the left- and right-moving sectors. We find
\be
	\beta_r = - \sqrt{3}\, \varphi_l\ , \qquad \omega^y_l = \pm i \pi\ , \qquad \beta_l = - \frac1{\sqrt{3}}\, \varphi_l\ ,
\ee
such that
\be
	I_r^\text{aBPS} = - \frac{2 \pi\, \varphi_r^3}{3 \sqrt{3}\,  ((\omega^x_r)^2+\pi^2) }  \ , \qquad \qquad I_l^\text{aBPS} = 0\ ,
\ee
with a quantum statistical relation for the right-movers,
\be
	I_r^\text{aBPS} = - S_r + 2\, \varphi_r\, Q - \omega_r^x\, J_x\ .
\ee
In order to restore the canonical normalization, we need to define
\be
	\varphi_\text{aBPS} : = -2\, \varphi_r\ , \qquad \omega^x_\text{aBPS} := \omega^x_r\ ,
\ee
such that
\be
\label{eq:5dabps}
	I_r^\text{aBPS} = - S_r - \varphi_\text{aBPS}\, Q - \omega_\text{aBPS}^x\, J_x = \frac{\pi\, \varphi_\text{aBPS}^3}{12 \sqrt{3}\,  ((\omega^x_\text{aBPS})^2+\pi^2) } \ .
\ee
Even though the answer is almost identical to the BPS case, as already commented on above the aBPS limit does not preserve supersymmetry for the same supergravity theory. Furthermore, the set of 4d conserved charges look drastically different from the ones in the BPS limit. The 4d BPS limit corresponds to a static black hole, since $J^y$ is the Kaluza-Klein electric charge $Q^0$. On the other hand, the aBPS limit includes the conserved charge $J^x$, which is the solely remaining rotation in 4d. Therefore the 4d black holes are indeed rotating, with angular momentum bounded from above by the absolute value of the electric charge Q. Furthermore, since these solutions only preserve supersymmetry in the presence of gauging, it turns out the OSV formula is not applicable and the attractor mechanism in this case is governed by the same fixed-point formula applicable to the asymptotically AdS black holes, see \cite{Hosseini:2019iad} and \cite{Hristov:2021qsw}. In this case supersymmetry actually fixes $\varphi^0_\text{aBPS} = -\pi\, P^0 = -\pi$, such that the fixed point formula of \cite{Hosseini:2019iad} gives
\be
	I^\text{fixed pt.}_{5d} =  (F_{5d} + \bar F_{5d}) \Big|_{(X^0 = \varphi^0_\text{aBPS} + i \omega_\text{aBPS} P^0; X^1 = \varphi^1_\text{aBPS})} =  \frac{2 \pi\, (\varphi^1_\text{aBPS})^3}{ ((\omega_\text{aBPS})^2+\pi^2) }\ ,
\ee
in agreement with \eqref{eq:5dabps} upon the identification $\omega = \omega^x$ and $\varphi^1 = \varphi/ (2 \sqrt{3})$ in the 4d and 5d pictures of the same physical system.

Note that the existence of an almost BPS limit is not a special property of black holes in 5d supergravity. Such solutions are generically present in matter-coupled 4d supergravity (but not in the minimal version), \cite{Goldstein:2008fq,Hristov:2012nu}, and in some form can be argued to exist in the asymptotically AdS black holes as well, see \cite{Gnecchi:2012kb}. We therefore expect the above structure to hold true in full generality, i.e.\ the right-moving sector to vanish in the BPS limit while the left-moving sector to vanish in a similar almost BPS limit.

\subsubsection*{Relation to the other extensions}
As already remarked in the previous sections, we expect the main features of 4d and 5d asymptotically flat black holes with extra matter and/or higher derivative corrections to remain largely the same, with varying degrees of technical challenges depending on the number of fields, conserved charges and respective couplings of the theory. The 5d model discussed in this section remarkably exhibits a higher degree of symmetry between the left- and right-moving sectors, which suggests that technically 5d models are actually simpler. The existence of two different special bounds, the BPS and the almost BPS one, is another feature that should remain true in general, but it appears technically harder to distinguish in the 4d matter-coupled theory. Interestingly, the addition of higher derivative corrections is expected to preserve the existence of the two different bounds, but the right-moving sector is supposed to receive many more HD corrections as opposed to the left-moving one in the respective BPS/aBPS limit. This has been discussed in some length in \cite{Hristov:2021qsw}, but the full HD contribution to the thermal black holes remains an open question. Likewise, it is interesting to consider minimal 5d {\it gauged} supergravity, exhibiting asymptotically AdS$_5$ black holes, \cite{Chong:2005hr}. These black holes exhibit three (doubled) horizons and are therefore conceptually distinct from the AdS$_4$ solutions discussed below. It remains a challenge to unearth a deeper structure in terms of left- and right-moving variables for such cases with an odd number of horizons, see also the asymptotically flat black holes in 7d {\it ungauged} supergravity, \cite{Cvetic:1996dt,Chow:2007ts}.

\section{AdS$_4$ asymptotics}
\label{sec:ads}

Now we consider 4d Einstein-Maxwell theory with a {\it negative} cosmological constant, 
\be
	 I_{\text{EM}-\Lambda} = \frac1{16 \pi} \int_{\cal M}  {\rm d}^4 x\, \sqrt{-g}\, \left( R + 6\, L^{-2}  - \frac14 F_{\mu \nu} F^{\mu \nu} \right)\ ,
\ee
with the constant $L$ setting the AdS$_4$ radius. This is the bosonic part of the action of 4d minimal {\it gauged} supergravity, which can be again defined from the prepotential
\be
	F_\text{min} (X) = - 2 i\, X^2\ ,
\ee
together with the gauging of the R-symmetry via the gauge field, $F = {\rm d} A$. Interestingly, asymptotically AdS$_4$ black holes are no longer restricted to only have spherical symmetry, see \cite{Caldarelli:1998hg}. Insisting on a compact horizon, we have a choice of a Riemann surface $\Sigma_\frak{g}$ horizon topology with any genus $\frak{g}$. In this work we further specialize to the cases of a sphere, $\frak{g} = 0$, with a positive metric curvature, and a higher genus case, $\frak{g} > 1$, corresponding to a discrete quotient of the hyperbolic space with a negative metric curvature. The most general Kerr-Newman-AdS$_4$ solution of this class, see \cite{Kostelecky:1995ei,Caldarelli:1998hg,Caldarelli:1999xj}, is given by the background metric~\footnote{Note that for the sake of simplicity here, we do  not discuss the so-called spindle horizons that do not correspond to smooth manifolds. One can find more about their thermodynamics in \cite{Cassani:2021dwa} and in principle repeat the following steps in an analogous way, but we leave this for future work.}
\be
	{\rm d} s^2 = - \frac{\Delta_r}{\Xi^2\, \rho^2} \left( {\rm d} t - a \sin^2 (\sqrt{\kappa}\, \theta)\, {\rm d} \phi \right)^2 + \frac{\rho^2}{\Delta_r}\, {\rm d} r^2 
+ \frac{\rho^2}{\Delta_\theta}\, {\rm d} \theta^2 +  \frac{\kappa\, \Delta_\theta\, \sin^2 (\sqrt{\kappa}\,\theta)}{\Xi^2\, \rho^2} \left( a\, {\rm d} t - (r^2+a^2)\, {\rm d} \phi \right)^2\ ,
\ee
with 
\bea
\begin{split}
	\rho^2 & = r^2 + a^2 \cos^2 \theta\ , \qquad \qquad \qquad \Xi = 1 -\kappa\, a^2\, L^{-2}\ , \\
	\Delta_r &=( r^2+ a^2)\, ( \kappa+L^{-2}\, r^2) - 2 m r  + q^2 + p^2\ , \qquad \Delta_\theta = 1 - \kappa\, a^2\, L^{-2}\, \cos^2 (\sqrt{\kappa}\,\theta)\ ,
\end{split}
\eea
where $\kappa = 1$ for the spherical topology and $\kappa = -1$ for the hyperbolic horizon. Note that we can only compactify the hyperbolic horizon to an arbitrary Riemann surface of genus $\frak{g} > 1$ only in the absence of rotation, $a = 0$. We further have the background gauge field
\be
	A = - \frac{q\, r}{\Xi\, \rho^2}\,  \left( {\rm d} t - a \sin^2 (\sqrt{\kappa}\, \theta)\, {\rm d} \phi \right) -  \frac{p\, \cos (\sqrt{\kappa}\, \theta)}{\Xi\, \rho^2}\,   \left( a\, {\rm d} t - (r^2+a^2)\, {\rm d} \phi \right)\ .
\ee
The conserved mass, $M$, angular momentum, $J$, and electromagnetic charges, $Q$ and $P$, are then given by
\be
	M = \frac{m}{\Xi^2}\, |1-\frak{g}|\ , \qquad J = \frac{a\, m}{\Xi^2}\, |1-\frak{g}|\ , \qquad Q = \frac{q}{\Xi}\, |1-\frak{g}|\ , \qquad P = \frac{p}{\Xi}\, |1-\frak{g}|\ ,
\ee
presented in a single formula by a slight abuse of notation, as one must remember that $\frak{g} > 1$ corresponds to $\Xi = 1$.

Looking at the metric component $g^{rr}$, there are now {\it four} independent roots of the function $\Delta_r$. We can denote them with $r_{(i)}$, $i = 1, 2, 3, 4$, noting that only two of them are actually {\it real} for the physical thermal black holes. We choose as a convention that the well-defined black hole range corresponds to real $r_3$ and $r_4$, with $r_3 \leq r_4$ and extremality reached when $r_3 = r_4$. In this regime the values of $r_1$ and $r_2$ remain complex, and conjugate to each other. Even though the explicit expressions for each $r_{(i)}$ are not illuminating (see the {\it .nb} file), it is interesting to note that the sum of four horizons actually vanishes identically,
\be
	\sum_{i=1}^4 r_{(i)} = 0\ .
\ee 
This is a direct consequence of Vieta's formula for the sum of the roots of a quartic polynomial.

We can define consistent thermodynamic variables on each of these horizons. The corresponding chemical potentials are, see \cite{Caldarelli:1999xj}, 
\bea
\label{eq:beta}
\begin{split}
	\beta_{(i)} & = 2 \pi\, \frac{r_{(i)}^2 + a^2}{\left( 2 L^{-2}\, r^3_{(i)}+r_{(i)}\, (\kappa+a^2\, L^{-2}) - m\right)}\ ,  \\
	\Phi_{(i)} &= \frac{q\, r_{(i)}}{r_{(i)}^2 + a^2}\ , \qquad \Psi_{(i)} = \frac{p\, r_{(i)}}{r_{(i)}^2 + a^2}\ , \qquad  \Omega_{(i)} = \frac{(1+a^2\, L^{-2})\, a}{r_{(i)}^2 + a^2}\ ,
\end{split}
\eea
and the Bekenstein-Hawking entropies are given by
\be
\label{eq:entropy}
	S_{(i)} = \frac{\pi}{\Xi}\, (r_{(i)}^2 + a^2)\, |1-\frak{g}|\ .
\ee
All of the chemical potentials obey the first law of thermodynamics,
\be
	\beta_{(i)}\, \delta M = \delta S_{(i)} + \beta_{(i)} \Omega_{(i)}\, \delta J + \beta_{(i)} \Phi_{(i)}\, \delta Q\ ,
\ee
and the quantum statistical relation
\be
	I_{(i)} = \beta_{(i)}\, M - S_{(i)}-\beta_{(i)} \Omega_{(i)}\, J - \beta_{(i)} \Phi_{(i)}\, Q\ . 
\ee
Note that here we are in the canonical potential for the magnetic charge, i.e.\ fixing $P$ rather than the conjugate $\Psi_{(i)}$, following the standard boundary conditions in holography, \cite{Hawking:1995ap}.

The main complication arising here to defining the left- and right-moving variables is the following: there are now various ways of generalizing the construction. If we insist on the variables to be based on two horizons only, we can define such a system of variables for {\it every} two pairs of horizons, $r_{(i)}$ and $r_{(j)}$:
\bea
\label{eq:pairwise}
\begin{split}
	\beta^{(ij)}_{l,r} :=& \frac12\, (\beta_{(i)} \pm \beta_{(j)})\ , \qquad \qquad \qquad \omega_{l,r}^{(ij)} :=  \frac12\, (\beta_{(i)} \Omega_{(i)} \pm \beta_{(j)} \Omega_{(j)})\ , \\  \varphi^{(ij)}_{l,r} :=& \frac12\, (\beta_{(i)} \Phi_{(i)} \pm \beta_{(j)} \Phi_{(j)})\ , \qquad \qquad S^{(ij)}_{l,r} :=  \frac12\, (S_{(i)} \pm S_{(j)})\ ,
\end{split}
\eea
and
\be
\label{eq:newvar2}
	I^{(ij)}_{l,r} := \frac12\, ( I_{(i)} \pm I_{(j)})\ ,
\ee
leading to the new first law,
\be
	\beta^{(ij)}_{l,r}\, \delta M = \delta S^{(ij)}_{l,r} + \omega_{l,r}^{(ij)}\, \delta J + \varphi^{(ij)}_{l,r}\, \delta Q\ ,
\ee
and corresponding quantum statistical relation,
\be
	I^{(ij)}_{l,r} = \beta^{(ij)}_{l,r} M - S^{(ij)}_{l,r} - \omega^{(ij)}_{l,r} J - \varphi^{(ij)}_{l,r} Q\ .
\ee

Apart from the above construction, the existence of four horizons gives us a much larger set of possible linear combinations, e.g.\ using all four horizons together. After some experimentation about this, with the help of {\it Mathematica} and the supplemented file, we have isolated a set of particularly interesting combinations of the four sets of variables:
\bea
\label{eq:directional}
\begin{split}
	\beta_\text{W} := \frac12\, (\beta^{(12)}_l - \beta^{(34)}_l) = \frac14\, \left(\beta_{(1)} + \beta_{(2)} - \beta_{(3)} - \beta_{(4)} \right)\ , \\ 
  	\beta_\text{N} := \frac12\, (\beta^{(12)}_r - \beta^{(34)}_r) = \frac14\, \left(\beta_{(1)} - \beta_{(2)} - \beta_{(3)} + \beta_{(4)} \right)\ , \\ 
	\beta_\text{E} := \frac12\, (\beta^{(12)}_l + \beta^{(34)}_l) = \frac14\, \left(\beta_{(1)} + \beta_{(2)} + \beta_{(3)} + \beta_{(4)} \right)\ , \\ 
	\beta_\text{S} := \frac12\, (\beta^{(12)}_r + \beta^{(34)}_r) = \frac14\, \left(\beta_{(1)} - \beta_{(2)} + \beta_{(3)} - \beta_{(4)} \right)\ , 
\end{split}
\eea
and analogous definitions for all other chemical potentials, entropies and on-shell actions. Naturally this also gives an automatically satisfied first law of black hole thermodynamics and quantum statistical relation for every on-shell action, $I_\text{W, N, E, S}$. We would then for example recover the on-shell action of the outer-most horizon by
\be
	I_{(4)} = I_\text{W} + I_\text{N} + I_\text{E} + I_\text{S}\ .
\ee

It seems natural to expect that the left- and right-moving sectors related to the outer two horizons, $I_{l, r}^{(34)}$, are the best candidate for interesting simplifications, as also considered in \cite{Nian:2020qsk} in the near-extremal limit of the AdS$_5$ solutions. However, it actually turns out that the latter ``directional'' parametrization seems most interesting from the point of view of the resulting expressions and the BPS limits. Let us first emphasize that no matter which variables we use, the explicit expressions are too complicated for a complete simplification and an explicit result, as can be verified by numerical checks in the {\it .nb} file. However, we find one remarkable simplification for the on-shell action in the latter coordinates, \eqref{eq:directional}, which is again a result of numerical analysis. Without taking any limits in the full phase space, we find a constant expression for $I_\text{E}$ that is independent of the related chemical potentials $\beta_\text{E}, \varphi_\text{E}, \omega_\text{E}$, and only depends by the sign of the horizon curvature and the respective genus,
\be
\label{eq:interestingsum}
	I_\text{E} = \frac{\kappa\, \pi}{2}\, L^2\, |1-\frak{g}|\ .
\ee
Note that the corresponding East-moving chemical potentials all vanish identically,
\be
	\beta_\text{E} = \varphi_\text{E} = \omega_\text{E} = 0\ ,
\ee
and the sum of the entropies then automatically obeys the above sum (with an overall minus sign due to the quantum statistical relation), as already noticed in \cite{Xu:2013zpa,Xu:2014qaa}. Ultimately these completely simplifications can be attributed to Vieta's formula for the sum of the roots $r_{(i)}$. The other three on-shell actions,  $I_\text{W}$,  $I_\text{N}$, and  $I_\text{S}$, are not constant in general and we could not infer their explicit dependence on the corresponding potentials. It would be very interesting to derive the above simple formula either from first principles or microscopically from a dual field theory point of view.

 We now move to discuss in more detail the corresponding BPS limits, which are rather distinct for the two main choices of horizon topology, and will be discussed separately. 

\subsection{Higher-genus topology}

In the higher-genus case,  $\kappa = -1$ and $\frak{g} > 1$, supersymmetry is realized via the so-called topological twist, see \cite{Romans:1991nq,Caldarelli:1998hg}. The BPS limit follows from the corresponding twisted superalgebra, see \cite{Caldarelli:1998hg,Hristov:2011ye,Hristov:2013spa}, and fixes the mass and magnetic charge in the following way:
\be
	M = m =  0\ , \qquad \qquad p = \frac{L}{2}\ .
\ee
As already explained above, higher-genus topology is only allowed for a vanishing angular momentum, $J = 0$, such that the only free asymptotic charge remains the electric one. This BPS limit, from the point of view of the procedure introduced in \cite{Cabo-Bizet:2018ehj,Cassani:2019mms}, was recently analyzed in \cite{BenettiGenolini:2023ucp}. Looking at the chemical potential on all four horizons under the above constraints, we find
\be
\label{eq:horizonsconstraints}
	2 L^{-1}\, \beta_{(1,2)}\, \Phi_{(1,2)}  = \pm (s_1)\, 2 \pi i\ , \qquad  2 L^{-1}\, \beta_{(3,4)}\, \Phi_{(3,4)} = \pm (s_2)\, 2 \pi i\ ,
\ee
with two unrelated arbitrary signs, $s_{1,2} = \pm 1$, which for the outer horizon was already noticed in \cite{BenettiGenolini:2023ucp}. In this case we can numerically verify that the on-shell action on {\it all} four horizons actually coincide,
\be
	I^\text{BPS}_{(1)} = I^\text{BPS}_{(2)} = I^\text{BPS}_{(3)} = I^\text{BPS}_{(4)} = - \frac{\pi}{2}\, L^2\, |1-\frak{g}|\ .
\ee

In this case we see that  both the pairwise construction on the outer two horizons, \eqref{eq:pairwise}, and the ``directional'' variables, \eqref{eq:directional}, reproduce the same answer that matches the results of \cite{BenettiGenolini:2023ucp}. Let us first look at the standard two outer horizons, $r_{(3, 4)}$, where we find from the above relations that
\be
	I_l^{(34), \text{BPS}} = - \frac{ \pi}{2}\, L^2\, |1-\frak{g}|\ , \qquad I_r^{(34), \text{BPS}} = 0\ ,
\ee
together with
\be
	L^{-1}\, \varphi_l^{(34)} = 0\ , \qquad \qquad L^{-1}\, \varphi_r^{(34)} = \pm 2 \pi i\ ,
\ee
and a quantum statistical relation
\be
	I^{(34), \text{BPS}}_l =  - S_l^{(34)}\ .
\ee
This is in agreement with the constant positive entopy for the BPS higher-genus black holes, which can also be understood as part of the general class of twisted black holes in matter-coupled gauged supergravities, see \cite{Benini:2015eyy,Benini:2016hjo,Azzurli:2017kxo}.

Looking at the ``directional'' chemical potentials, we also find
\be
	L^{-1}\, \varphi_\text{W} = L^{-1}\, \varphi_\text{E} = 0\ ,
\ee
and depending on the sign of  $s_{1,2}$ in \eqref{eq:horizonsconstraints}, either
\be
	L^{-1}\, \varphi_\text{N} = 0\ , \qquad \qquad L^{-1}\, \varphi_\text{S} = \pm 2 \pi i\ ,
\ee
or the other way around, $\text{N} \leftrightarrow \text{S}$. Furthermore, we simply have
\be
	I^\text{BPS}_\text{N} = I^\text{BPS}_\text{W} = I^\text{BPS}_\text{S} = 0\ , \qquad \qquad I^\text{BPS}_\text{E} = - \frac{ \pi}{2}\, L^2\, |1-\frak{g}|\ ,
\ee
and a quantum statistical relation
\be
	I^\text{BPS}_\text{E} = - S_\text{N}\ .
\ee
Clearly, the two sets of variables both seem to be equally adequate in reconstructing correctly the BPS limit of the higher-genus black holes. We turn next to the spherical solutions and their corresponding BPS bound.

\subsection{Spherical topology with rotation}

In the spherical case,  $\kappa = 1$ and $\frak{g} = 0$, supersymmetry is realized without any twist, leaving the spinors non-constant on the horizon, see \cite{Kostelecky:1995ei,Caldarelli:1998hg,Hristov:2011ye}. In this case the angular momentum is generically non-zero, and instead the mass and magnetic charge are fixed as follows:
\be
	M = L^{-1}\, |J| + |Q|\ , \qquad \qquad P = 0\ ,
\ee
in which case we can numerically verify the relations
\bea
\begin{split}
	\beta_{(1,2)}\, \left( L^{-1} + \Omega_{(1,2)} - 2 L^{-1}\, \Phi_{(1,2)} \right) & = \pm (s_1)\, 2 \pi i\ , \\
	\beta_{(3,4)}\, \left( L^{-1} + \Omega_{(3,4)} - 2 L^{-1}\, \Phi_{(3,4)} \right) & = \pm (s_2)\, 2 \pi i\ ,
\end{split}
\eea
on the four event horizons, respectively, with two unrelated arbitrary signs, $s_{1,2} = \pm 1$. This relation for the outer horizon was already observed in \cite{Cassani:2019mms}.

When translated to the ``directional'' variables, these relations lead to
\be
	L^{-1}\, \beta_\text{W} + \omega_\text{W}-2 L^{-1}\, \varphi_\text{W} = L^{-1}\, \beta_\text{E} + \omega_\text{E}-2 L^{-1}\, \varphi_\text{E} = 0\ ,
\ee
and, depending on the sign of  $s_{1,2}$, either
\be
\label{eq:corr}
	 \beta_\text{N} + \omega_\text{N}-2 L^{-1}\, \varphi_\text{N} = 0\ , \qquad  \beta_\text{S} + \omega_\text{S}-2 L^{-1}\, \varphi_\text{S}  = \pm 2 \pi i\ ,
\ee
or the other way around, $\text{N} \leftrightarrow \text{S}$. Similarly one can also find the respective relations in the pairwise left- and right-moving variables, \eqref{eq:pairwise}. 

Unfortunately, the present BPS limit turns out to be substantially more complicated to simplify, and numerically we could only verify the interesting relations,
\be
	I^\text{BPS}_{(1)} = (I^\text{BPS}_{(2)})^*\ , \qquad I^\text{BPS}_{(3)} = (I^\text{BPS}_{(4)})^*\ ,
\ee
with only their sum being fixed as in \eqref{eq:interestingsum}. In this case focusing only on the two outer horizons does not seem to give any clear simplification in the BPS limit, since
\be
	I_{l}^{(34), \text{BPS}} \neq I_{r}^{(34), \text{BPS}} \neq 0\ .
\ee
On the other hand, we do find a suggestive simplification for the ``directional'' on-shell actions, with either 
\be
	I^\text{BPS}_\text{N} = 0\ , \qquad  I^\text{BPS}_\text{W} \neq I^\text{BPS}_\text{S} \neq 0\ , \qquad I^\text{BPS}_\text{E} = \frac{ \pi}{2}\, L^2\ ,
\ee
or the exchange $\text{N} \leftrightarrow \text{S}$ in correlation with \eqref{eq:corr}. Furthermore, $I^\text{BPS}_\text{W}$ is purely real, while $I^\text{BPS}_\text{S}$ (or respectively $I^\text{BPS}_\text{N}$) is purely imaginary. Due to the limited power of the numerical analysis, we were not able to extract the dependence of the non-zero on-shell actions above in terms of the respective chemical potentials. This also means we could not derive an obvious map between the BPS actions in the natural variables suggested above and the result of \cite{Cassani:2019mms}, which agrees with the original observation of \cite{Choi:2018fdc} and the Cardy limit of the dual superconformal index, \cite{Choi:2019dfu}. Also in this case that the BPS on-shell action can be understood from a fixed-point formula, see \cite{Hosseini:2019iad}, and corresponds to
\be
	I^\text{fixed pt.}_{AdS_4} =  \frac{1}{4\, \omega_\text{BPS}}\, F_{min} \Big|_{(X = \varphi_\text{BPS})} = - \frac{i\, (\varphi_\text{BPS})^2}{2\, \omega_\text{BPS}}\ ,
\ee
under the constraint $2 L^{-1}\, \varphi_\text{BPS} - \omega_\text{BPS} = \mp 2 \pi i$, see \cite{Cassani:2019mms}. Interestingly, picking the upper sign we find explicitly
\be
	I^\text{fixed pt.}_{AdS_4} =  \frac{ \pi}{2}\, L^2 + \frac{i \pi^2\, L^2}{2\, \omega_\text{BPS}} - \frac{i\, L^2\, \omega_\text{BPS}}{8}\ , 
\ee
for an unconstrained $\omega_\text{BPS}$, such that the first term is in agreement with $ I^\text{BPS}_\text{E}$ above, and the other two terms should match $ I^\text{BPS}_\text{W} + I^\text{BPS}_\text{S}$. It remains an open question to fully derive this result in the approach of the natural variables. On the other hand, even without this explicit map, it is clear that the BPS limit leads to simplifications for the ``directional'' variables, and it is an interesting question how to interpret these results holographically.

\subsubsection*{Relation to the other extensions}
From the above discussion it is clear that the inclusion of a cosmological constant presents a rather formidable technical complication, which is not always amenable even to numerical analysis. Therefore we expect the further superposition of additional generalizations to only worsen the situation. As we already commented, the addition of four derivative terms to minimal gauged supergravity is actually under complete control, see \cite{Bobev:2020egg,Bobev:2021oku}, such that similarly to \eqref{eq:4dlrsectors} only the left-moving, or ``E'' direction, will get shifted by a constant term. On the other hand, the inclusion of other HD terms, any form of matter, and/or the generalization to 5d with AdS asymptotics seem much more complicated (see again \cite{Nian:2020qsk} for results in the latter direction).

\section*{Acknowledgements}
I am very grateful to S.\ M.\ Hosseini and R.\ G.\ Pozzi for helpful discussions. This study is financed by the European Union- NextGenerationEU, through the National Recovery and Resilience Plan of the Republic of Bulgaria, project No BG-RRP-2.004-0008-C01.

\bibliographystyle{JHEP}
\bibliography{newthermo.bib}

\end{document}